\documentclass[epj]{webofc}
\usepackage[varg]{txfonts}   % Web of Conferences font
\wocname{EPJ Web of Conferences}
\woctitle{New Frontiers in Physics 2014}
\begin{document}
\selectlanguage{english}
\title{Testing General Relativity and Alternative Theories of Gravity with Space-based Atomic Clocks and Atom Interferometers}
%
% subtitle is optional
%
%%%\subtitle{Do you have a subtitle?\\ If so, write it here}

\author{Ruxandra Bondarescu \inst{1}\fnsep\thanks{\email{ruxandra@physik.uzh.ch}} \and
        Andreas Sch\"arer, \inst{1} 
        Philippe Jetzer\inst{1} \and
        Raymond Ang\'elil \inst{2} \and
        Prasenjit Saha \inst{2} \and
        Andrew Lundgren\inst{3}
        % etc.
}

\institute{Department of Physics, University of Z\"urich, Z\"urich, Switzerland
\and
           Institute for Computational Science, University of Z\"urich, Z\"urich, Switzerland 
           \and Albert Einstein Institute, Hannover, Germany
          }

\abstract{The successful miniaturisation of extremely accurate atomic clocks and atom interferometers invites prospects for satellite missions to perform precision experiments. We  discuss the effects predicted by general relativity and alternative theories of gravity that can be detected by a clock, which orbits the Earth. Our experiment relies on the precise tracking of the spacecraft using its observed tick-rate. The spacecraft's reconstructed four-dimensional trajectory will reveal the nature of gravitational perturbations in Earth's gravitational field, potentially differentiating between different theories of gravity. This mission can measure multiple relativistic effects all during the course of a single experiment, and constrain the Parametrized Post-Newtonian Parameters around the Earth. A satellite carrying a clock of fractional timing inaccuracy of $\Delta f/f \sim 10^{-16}$ in an elliptic orbit around the Earth would constrain the PPN parameters $|\beta -1|, |\gamma-1| \lesssim 10^{-6}$.  We also briefly review potential constraints by atom interferometers on scalar tensor theories and
in particular on Chameleon and dilaton models.}
\maketitle
\section{Introduction}
\label{intro}
\begin{figure}[htb]
%% Use the relevant command for your figure-insertion program
%% to insert the figure file.
\centering
\sidecaption
\includegraphics[width=7cm,clip]{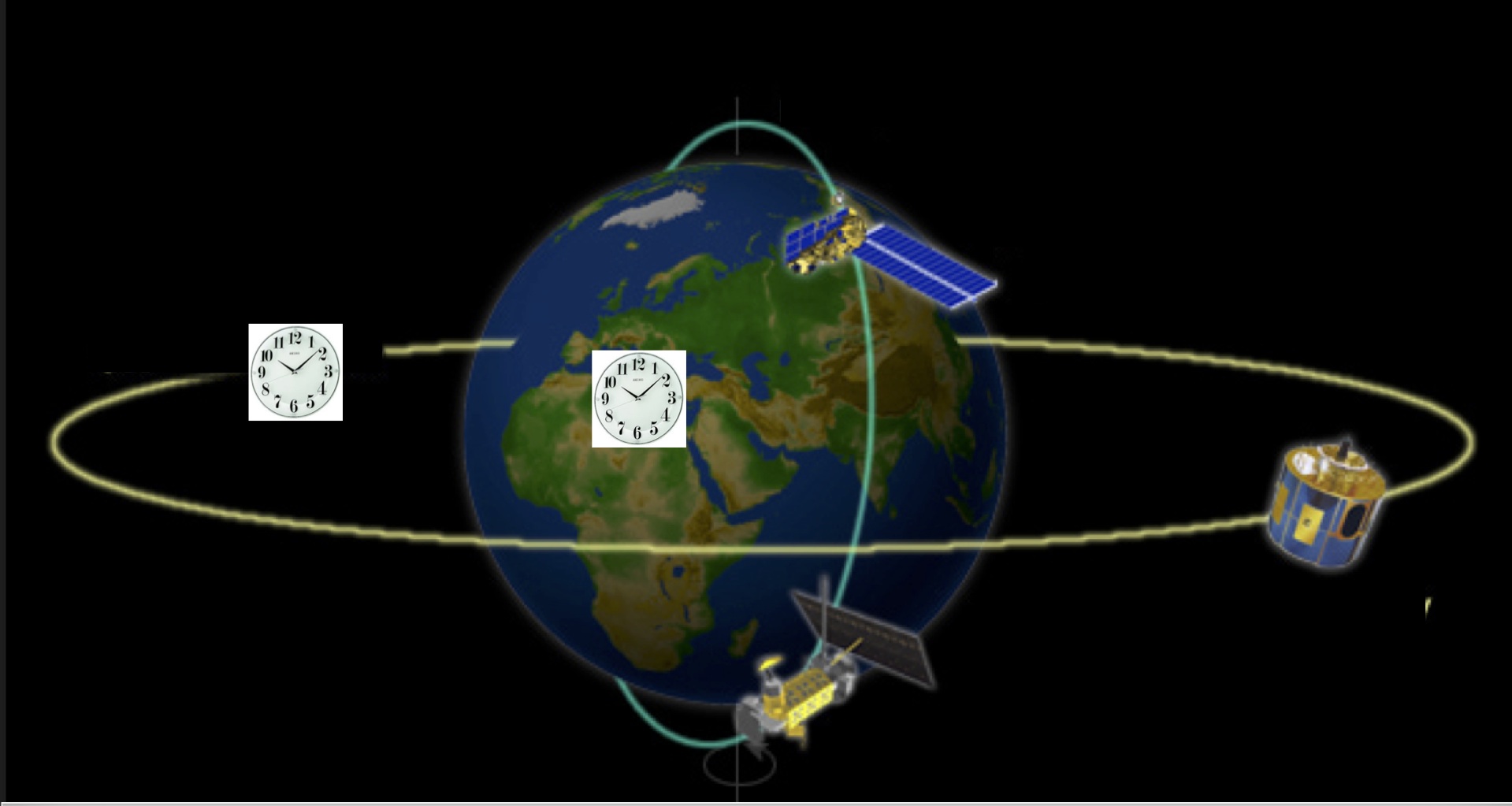}
\caption{An atomic clock in an elliptic orbit travelling around the Earth together with other proposed satellites.}
\label{fig-1}       % Give a unique label
\end{figure}
There are a number of proposed missions that plan to carry atomic clocks or atom interferometers in space. The Atomic Clock Ensemble 
in Space (ACES) will place two ultra-precise atomic clocks on the International Space Station in mid-2016
that communicate with clocks at receiving stations on Earth with the aim to reach a fractional timing inaccuracy of  $\Delta f/f \sim 10^{-16}$ \cite{aces}.
This mission will be followed by the Space Optical Clock (SOC), which is expected to place the first optical clock
on the ISS. Another potential mission that may link clocks in space with the best available clocks on Earth is the Gravitational Redshift explorer (GRESE)
\cite{GRESEWeb}, which is a proposed joined mission between China and Switzerland that aims to place a clock in an elliptical orbit around Earth. In addition to these,  the Space-Time 
Explorer and Quantum Equivalence Test (STE-QUEST) satellite proposes to place an atom interferometer in Earth orbit, which can perform
tests of the Universality of Free Fall (UFF) at the quantum level \cite{sq}, and the MICROSCOPE experiment is already approved to test the UFF
 with two one kilogram macroscopic masses \cite{microscope}.

We discuss the ability of these missions to constrain (1) higher order relativistic effects (Sec. 3), (2) alternative theories of gravity 
via parametrized Post-Newtonian parameters (Sec. 4), and (3) the universality of free-fall through the E\"{o}tv\"{o}s parameter (Sec. \ref{Uff}).
This proceeding reviews parts of \cite{Ray, Andreas,masterthesis} and also describes some additional results.

\section{Brief Overview of Atomic Clocks and Atom Interferometers}
Progress in atomic clock technology already has had tremendous impact on our every day life.
The frequency stability of atomic clocks has been improving at a rate of about a factor of 10 every
decade for the past 60 years. After the discovery of the femtosecond laser frequency comb, which enabled
the counting of the oscillations of optical atomic transition, optical clocks have been
 improving at an even more rapid rate \cite{Poli2013}. Clocks are used to define both the meter and the second. 
Since 1983, the meter has been defined as the length of the path travelled by light in a vacuum
in $1/299 792 458$ of a second. This means that already in 1983 we were better at measuring 
1 part in $10^9$ of a second than at measuring the length of an object. 

The best atomic clocks on Earth demonstrate clock stability of $\Delta f/f\approx 3 \times 10^{-16}/\sqrt{\tau/\rm{sec}}$ for averaging
 times $\tau$. They reach a frequency stability of  $\Delta f/f \approx 1.6 \times 10^{-18}$ in an averaging time of 25, 000 s 
 or about 7 hours \cite{Hinkley2013}. Optical transfer over free space has achieved to an accuracy of $10^{-18}$ over a 
 distance of a few kilometers \cite{Giorgetta}.

%Optical clocks, the most accurate atomic clocks, are not yet compact or reliable enough to be flown into space. The most
% stable space-qualified clock is a Cesium fountain clock. It has already been built and is expected to reach a precision 
% $\Delta f/f \sim 10^{-16}$ when placed on the International Space Station in 2016 as part of the ACES mission.  CSEM
%  in Neuch\^{a}tel developed and tested the space-qualified Hydrogen Maser clock for the ACES mission, which is the
%   second clock of the mission, and prepares to develop instrumentation for more futuristic missions like the STE-QUEST. 

The clocks in space will communicate with some of the best available clocks on Earth that are placed in strategic locations 
so that the satellite can always perform common-view comparisons. In the future, it is likely that many missions will either 
carry their own clock or carry transponders that can act as a mirror amplifying and retransmitting tick signals from other
clocks for orbit determination.
%However, while planned and future space missions stimulate atomic clock technologies to 
%improve, they provide no direct funding for theoretical work.  
    
Comparisons between atomic clocks on Earth and atomic clocks in space and atomic clock networks on Earth could be used for
relativistic geodesy, and have potential in adding detail to satellite maps, and in improving our understanding of the interior of the Earth \cite{Bondarescu2012},
of the solid Earth tide, and of processes such as earthquakes and volcanoes \cite{otherProceeding}.

A differential atom interferometer compares the free propagation of matter waves of different composition under the effect of 
gravity. It compares the free fall of atoms of different composition. The aim of the STE-QUEST mission is to compare the free
fall of $^{87}$Rb and $^{41}$K. Atom interferometers on the ground have the potential to be used as quantum gravimeters, and
hence have a plethora of applications in geophysics \cite{otherProceeding}.

\section{Testing General Relativity with Clocks in Space}
\begin{figure}[htb]
%% Use the relevant command for your figure-insertion program
%% to insert the figure file.
\centering
%\sidecaption
%\includegraphics[width=7cm,clip]{satellites.jpg}
\includegraphics[width=7cm,clip]{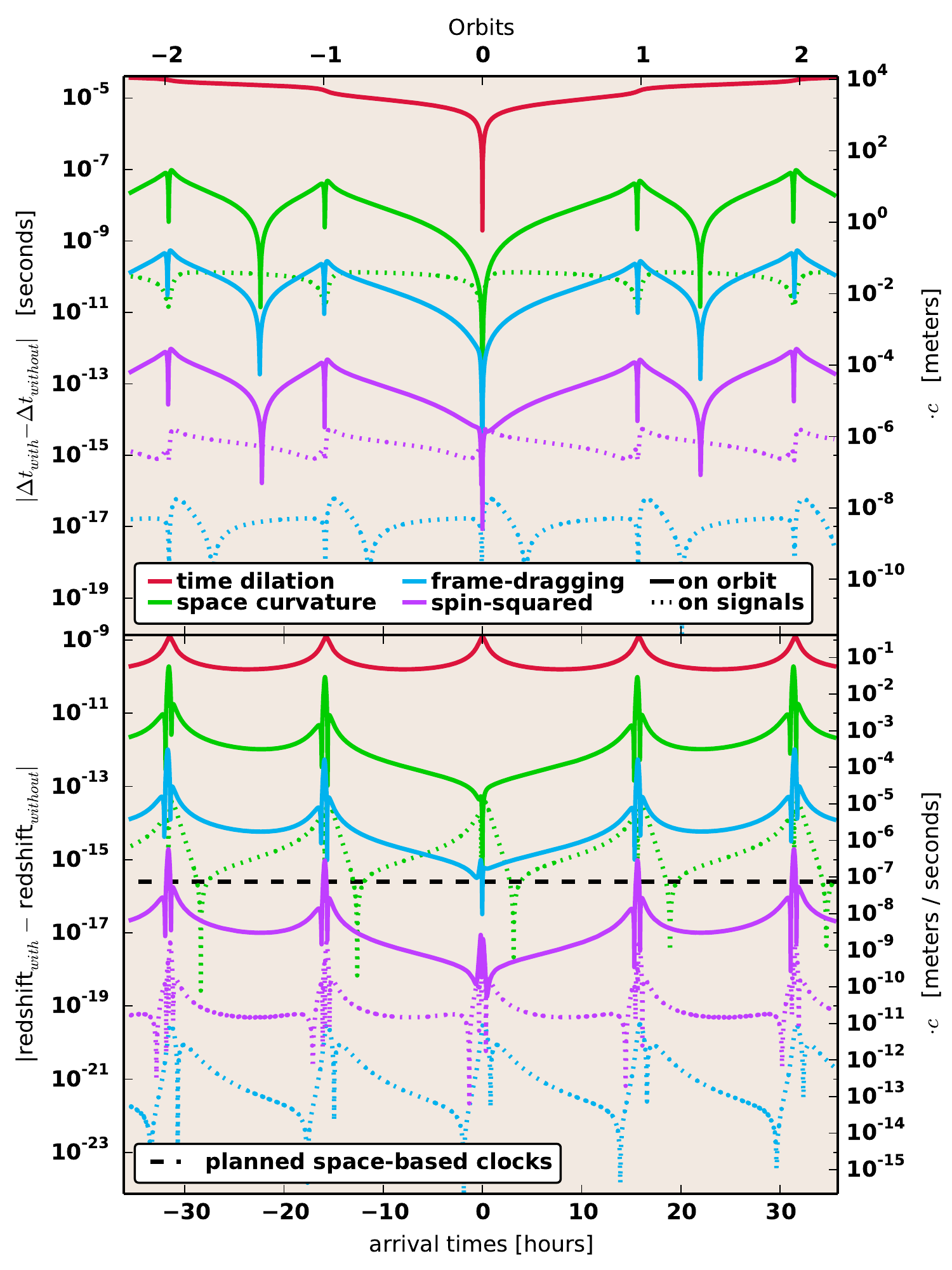}
\caption{Timing signals and the associated redshift signals from \cite{Ray} for a clock orbiting the Earth on the elliptical orbit originally
proposed for the STE-QUEST satellite, and now considered for the Redshift explorer.}
\label{fig-2}       % Give a unique label
\end{figure}

 We first present results from \cite{Ray} where we computed the orbit of the space-craft and the trajectory of the tick signals.
 A satellite broadcasts tick signals (light pulses) to an Earth-based receiving station,  whose arrival times are compared with a local clock. 
 Our numerical simulator calculates the orbit of the satellite and the trajectories of the broadcasted clock ticks. We found that clocks in space with a fractional
 timing inaccuracy of $\Delta f/f \sim 10^{-16}$ may probe a number of higher order relativistic effects. In particular:
 (1) Schwarzschild perturbations will be measurable through their effects both on the orbit and on the signal propagation, 
 (2) frame-dragging of the orbit will be readily measurable, and (3) in optimistic scenarios, the spin-squared metric effects 
 may be measurable for the first time ever.   One advantage of such an experiment would be that it would simultaneously probe different 
 relativistic effects with the same instrument in the same astrophysical system. Up until now, such effects were probed by various experiments 
  in independent astrophysical systems. For example, Shapiro delay has been measured in binary pulsars systems \cite{Lorimer}
  and frame dragging has been measured by Gravity Probe B \cite{Everitt} and LaRes \cite{Renzetti}, while spin squared 
  effects have yet to be measured.  Our computer program is freely available online \cite{Ray}.
 
 We note that we only solve the forward problem, which provides the first
 estimates for these effects without performing signal recovery. A fractional accuracy of $\Delta f/f \sim 10^{-16}$ 
 is expected to be reached by the space-qualified Hydrogen Maser clock built in Switzerland used in conjunction with a Caesium fountain clock. 
 Both will be placed on the International Space Station as part of ESA's ACES mission. Unfortunately, due to thrusting manoeuvres, 
 atmospheric drag and other ballistic accelerations the ISS is not freely falling and thus ACES and SOC cannot perform such a timing experiment, 
 but the Redshift explorer, if funded, will attempt to recover gravitational accelerations and may see these higher order general relativistic 
 effects (See Fig.\ \ref{fig-2}.).  
\section{Constraining Weak Field Gravity with Clocks in Space}
 \begin{figure}[htb]
%% Use the relevant command for your figure-insertion program
%% to insert the figure file.
\centering
%\sidecaption
%\includegraphics[width=7cm,clip]{satellites.jpg}
\includegraphics[width=13cm,clip]{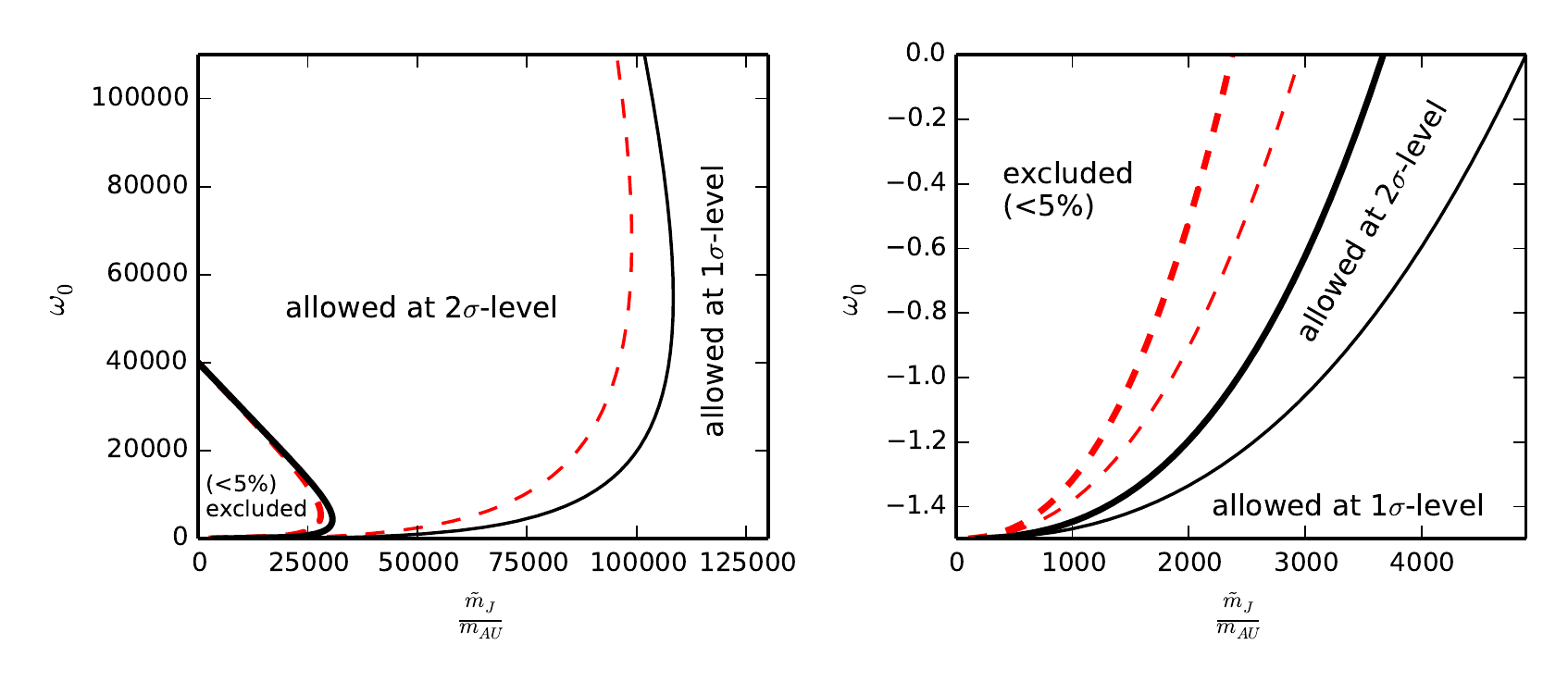}
\caption{Constraint on the $\gamma$-PPN parameter on massive Brans-Dicke theory from \cite{Andreas}. The solid lines show the 1-$\sigma$ and 2-$\sigma$
level constraints at 0.6 solar radii, and the dashed red lines show the corresponding result for the point mass approximation. }
\label{fig-3}       % Give a unique label
\end{figure}

\begin{figure}[htb]
%% Use the relevant command for your figure-insertion program
%% to insert the figure file.
\centering
%\sidecaption
%\includegraphics[width=7cm,clip]{satellites.jpg}
\includegraphics[width=7cm,clip]{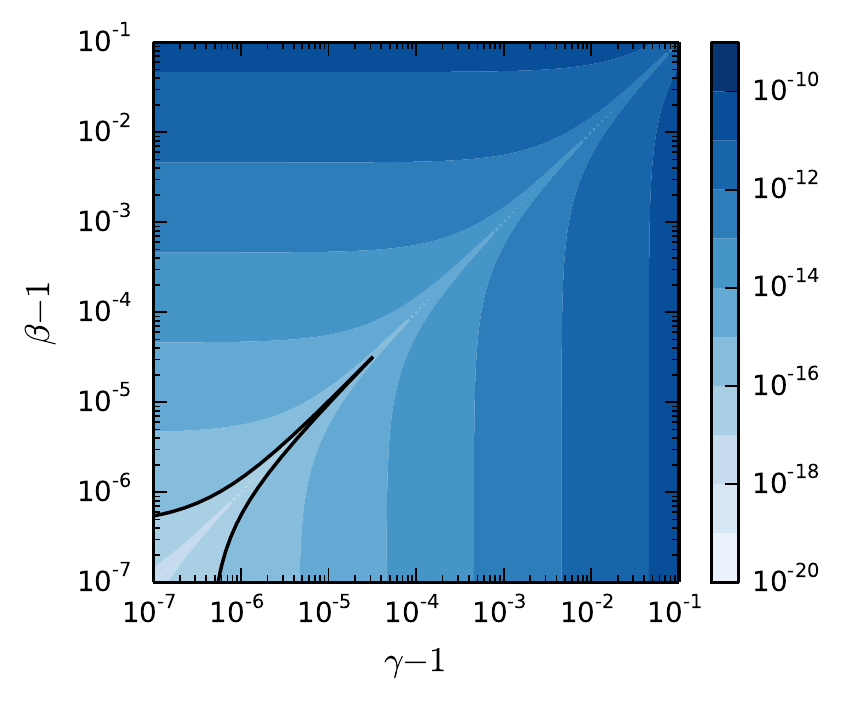}
\caption{Logarithmic plot of the difference in redshift between the GR-orbit and that for a range of
positive values of the $\gamma$ and $\beta$ PPN parameters over one orbit from \cite{Andreas}.
The solid line corresponds to $\Delta f/f \sim 10^{-16}.$}
\label{fig-4}       % Give a unique label
\end{figure}

 % both (1) the formalism necessary to Parametrized Post-Newtonian Framework (PPN) parameters for general scalar
%tensor theories \cite{Andreas}, as well as (2)
We further present potential constraints that could be placed on Parametrized Post-Newtonian Framework (PPN) parameters, which 
describe deviations from general relativity (GR) in the weak field regime, with an atomic clock in Earth orbit \cite{Andreas}. The PPN framework introduces a number of 
parameters in the various metric terms, which can vary with the location of the experiment. At present the strongest constrained
 PPN parameter is the $\gamma$ parameter. The Cassini space-craft predicts that $|\gamma - 1| < 2.3 \times 10^{-5}$ 
 (at the $1-\sigma$ level) around the Sun \cite{cassini}.
Planetary ephemerides are used to constrain $|\gamma - 1|$ and $|\beta - 1|$ to the $10^{-5}$ level \cite{beta}. 
GR itself predicts $\gamma=1$ and $\beta = 1$. We recompute the Cassini constraint for a massive Brans-Dicke scalar field
  inside and outside a constant density sphere and compare it with the point source result (See Fig.\ \ref{fig-3}.).
  
 In our recent paper \cite{Andreas} we computed the $\gamma$ and $\beta$ PPN parameters 
for general scalar-tensor theories formulated in the Einstein frame. We emphasize that it is natural for the PPN parameters to be environment dependent, 
and so it make sense to test them at different distances and in the gravitational fields of different objects \cite{Andreas}. We estimate the perturbations
caused by PPN parameters that deviate from GR on the trajectory of a space-craft carrying an
 atomic clock that orbits the Earth. Our estimates suggest that for a satellite on an eccentric orbit similar to that
 of the Redshift Explorer and to the originally proposed STE-QUEST mission deviations of the order of  $|\gamma - 1|, |\beta - 1| \sim 10^{-6}$ may be
 detectable for a clock with fractional frequency uncertainty $\Delta f/f \sim10^{-16}$ (See Fig.\ \ref{fig-4}.). Such space 
 experiments are within reach of the existing atomic clock technology. However, in order to make definite statements on detectability
 we would have to be able to recover such signals from realistic noise.

\section{Testing scalar-tensor theories with atom interferometers}
\label{Uff}
The weak equivalence principle states that two test masses of different compositions fall with the same acceleration in an external gravitational field. Therefore, by measuring the accelerations of two different bodies $a_1$ and $a_2$, one can calculate the E\"{o}tv\"{o}s parameter
\begin{align*}
\eta := 2 \frac{\left|a_1 - a_2\right|}{\left|a_1 + a_2\right|}.
\end{align*}
The best measurements of $\eta$ come from torsion-balance experiments, performed by the E\"{o}t-Wash group \cite{Torsion-balance tests of the weak equivalence principle 2012}. They constrain $\eta$ to be zero within an accuracy of about
\begin{align*}
\eta \lesssim 10^{-13}
\end{align*}
and of course a non-zero measurement of $\eta$ would indicate a deviation from general relativity. For more details see \cite{masterthesis}.

\subsection{Chameleon Models}
The chameleon mechanism introduced by Khoury and Weltman \cite{khoury} provides a possibility to accommodate a very light cosmological scalar field that couples to matter with gravitational strength to current observational constraints. The idea is that the field is nearly massless on cosmological scales such that it could act as quintessence to cause the accelerated expansion of the universe. But the field couples to matter fields in such a way that the mass of the scalar field depends on the surrounding matter density. In a dense region, like in and around the Earth, it becomes very massive such that the force corresponding to the scalar field becomes short range. This makes it very difficult to detect the chameleon particle by experiments performed on the Earth. Far away in space the effects due to the field would be much stronger. In fact, near future satellite missions could detect violations of the equivalence principle or an effective gravitational constant even at a level that has been excluded on Earth. This would be a strong indication that some hiding mechanism like the chameleon mechanism is present in nature.

We consider the Earth with mass $M_\oplus$ and radius $R_\oplus$ as a perfect sphere with homogeneous density surrounded by a background of homogeneous density. The coupling constant of the Earth to the chameleon field is $k_\oplus$  and if we assume the Earth to have a thin shell the approximate field profile outside is (\cite{khoury}, equ. 15)
\begin{align*}
&\phi_\oplus(r) \simeq -\frac{k_\oplus}{4 \pi M_{\text{Pl}}} 3 \frac{\Delta R_\oplus}{R_\oplus} \frac{M_\oplus}{r} + \phi_\infty
= -\frac{k_{\text{eff}\oplus}}{4 \pi M_{\text{Pl}}}  \frac{M_\oplus}{r} + \phi_\infty
\end{align*}
where the exponential term was dropped for simplicity and we defined the effective coupling of the Earth to the field $k_{\text{eff}\oplus} := 3 k_\oplus \frac{\Delta R_\oplus}{R_\oplus}$.

Further we consider a test particle somewhere out in space with mass $m_i$, radius $R_i$ and coupling constant to the chameleon field $k_i$ . The test particle is supposed not to have a thin shell such that its exterior field profile is (\cite{khoury}, equ. 18)
\begin{align*}
&\phi_i(r) \simeq -\frac{k_i}{4 \pi M_{\text{Pl}}} \frac{m_i}{r} + \phi_\infty.
\end{align*}

A test body with mass $m_i$ and coupling constant $k^{i}_\text{eff}$ exposed to an external scalar field $\phi$ experiences a force (\cite{khoury}, equ. 20)
\begin{align*}
\vec{F}^i_\phi = - \frac{k^{i}_\text{eff}}{M_\text{Pl}}m_i \vec{\nabla}\phi.
\end{align*}

Since our test mass in space does not have a thin shell we have $k_i = k^{i}_\text{eff}$ and plugging in the field profile outside the Earth gives
\begin{align*}
\vec{F}^i_\phi &= - \frac{k_i}{M_\text{Pl}}m_i \vec{\nabla}\phi
\simeq - \frac{k_i}{M_\text{Pl}}m_i \frac{k_{\text{eff}\oplus}}{4 \pi M_{\text{Pl}}} \frac{M_\oplus}{r^2} \hat{r}
= - \frac{2 k_i k_{\text{eff}\oplus}}{8 \pi M_\text{Pl}^2} \frac{m_i M_\oplus}{r^2} \hat{r}
\\
&= - 2 k_i k_{\text{eff}\oplus} \frac{G m_i M_\oplus}{r^2}  \hat{r}
= - 6 k_i k_\oplus \frac{\Delta R_\oplus}{R_\oplus} \frac{G m_i M_\oplus}{r^2} \hat{r}.
\end{align*}

The total force the mass experiences is the sum of the gravitational force ($F_\text{G}^i = G m_i M_\oplus / r^2$) and the chameleon force and thus its total acceleration is
\begin{align*}
a_i = a_\text{G}^i + a_\phi^i = \frac{F_\text{G}^i + F_\phi^i}{m_i}
= \frac{G M_\oplus}{r^2} \left(1 + 2 k_i k_{\text{eff}\oplus} \right).
\end{align*}

If we now consider two such test masses $m_1$ and $m_2$ the Etv\"{o}s parameter can be calculated
\begin{align*}
\eta &= 2\frac{|a_1 - a_2|}{|a_1 + a_2|}
= 2\frac{\left| \left(1 + 2 k_1 k_{\text{eff}\oplus} \right) - \left(1 + 2 k_2 k_{\text{eff}\oplus} \right) \right|}{\left| \left(1 + 2 k_1 k_{\text{eff}\oplus} \right) + \left(1 + 2 k_2 k_{\text{eff}\oplus} \right)  \right|}
\\
&= 2\frac{\left| k_{\text{eff}\oplus}\left(k_1 - k_2 \right) \right|}{\left| 1 + k_{\text{eff}\oplus}\left(k_1 + k_2 \right) \right|}
\\
&\simeq 2 \left|k_1 - k_2 \right| k_{\text{eff}\oplus}
\\
&= 6 \left|k_1 - k_2 \right| k_\oplus \frac{\Delta R_\oplus}{R_\oplus},
\end{align*}

where we expanded for small $k_{\text{eff}\oplus}$ which is reasonable since $k_\oplus,k_i \sim \mathcal{O}(1)$ and $\frac{\Delta R_\oplus}{R_\oplus} \ll 1$ due to the thin shell of the Earth.
\subsection{Dilaton Couplings}
In this subsection the framework of the dilaton scalar field, leading to violations of the weak equivalence principle, is reviewed. Here we will follow the notation of \cite{Damour.Donoghue}.

It is assumed that the field couples to an effective Lagrangian which contains three particles, the electron e as well as the u and the d quarks, and the two interactions due to the electromagnetic field $A_\mu$ and the gluonic field $A_\mu^\text{A}$. The basic idea is that the dilaton field couples with different strength to these five components. Therefore the coupling of the field to the mass terms of these three particles and to the two gauge fields is characterized by the five dimensionless dilaton-coupling parameters $d_{m_\text{e}}$ (electron), $d_{m_\text{u}}$ (u quark), $d_{m_\text{d}}$ (d quark), $d_{\text{e}}$ (electromagnetic field) and $d_{\text{g}}$ (gluonic field). The coupling to the two quark mass terms is redefined by $d_{\hat{m}}$ and $d_{\delta m}$ corresponding to $\hat{m} := \frac{1}{2}(m_\text{d} + m_\text{u})$ and $\delta m := m_\text{d} - m_\text{u}$, respectively. It turns out that demanding that the field couples differently to these five components implies that there are five dimensionless `constants' of nature which depend on the scalar field. In particular these are the fine-structure constant $\alpha = \frac{e^2}{4\pi}$, the (dimensionless) particle masses $\kappa m_\text{e}$, $\kappa m_\text{u}$, $\kappa m_\text{d}$ and $\kappa \Lambda_3$. Here $\kappa$ is the inverse Planck mass $\kappa = \sqrt{4\pi G}$ and $\Lambda_3$ is the QCD (quantum chromodynamics) energy scale. These fundamental constants will be denoted $k_\text{a}$, where $\text{a}$ is the index corresponding to one of the constants.

Also here the scalar field $\phi$ is assumed to have units of mass, where it is also useful to introduce the dimensionless dilaton field $\varphi := \kappa \phi$. Then the coupling strength of a mass made of species A, $m_\text{A}$, is given by \cite{Damour.Piazza}
\begin{align*}
\alpha_\text{A} &= \partial_{\kappa \phi} \ln \left( \kappa m_\text{A}(\phi) \right) = \partial_\varphi \ln \left( \kappa m_\text{A}(\varphi) \right)
\\
&= \sum_\text{a} \partial_{k_\text{a}} \ln \left( \kappa m_\text{A}(\varphi) \right) \frac{\partial k_\text{a}}{\partial \varphi}.
\\
&= \sum_\text{a} \frac{\partial \ln \left( \kappa m_\text{A}(\varphi) \right)}{\partial \ln k_\text{a}} \frac{\partial \ln k_\text{a}}{\partial \varphi}.
\end{align*}
Here we used the fact that the five fundamental constants $k_\text{a}$ depend on the scalar field, which allowed us to apply the chain rule. In fact, the coupling constants for the individual components of the effective Lagrangian are precisely $d_\text{a} = \frac{\partial \ln k_\text{a}}{\partial \varphi}$. Further we can define dilaton charges
\begin{align*}
Q^\text{A}_{k_\text{a}} := \frac{\partial \ln \left( \kappa m_\text{A}(\varphi) \right)}{\partial \ln k_\text{a}}
\end{align*}
which can be calculated for all particles and interactions. Again taking $\hat{m}$ and $\delta m$ instead of $m_\text{u}$ and $m_\text{d}$, the dilaton charges turn out to be \cite{Damour.Donoghue}
\begin{align*}
&Q_{\hat{m}} = F_\text{A} \left[ 0.093 - \frac{0.036}{A^{1/3}} - 0.020 \frac{(A-2Z)^2}{A^2} - 1.4 \cdot 10^{-4} \frac{Z(Z-1)}{A^{4/3}} \right]
\\
&Q_{\delta m} = F_\text{A} \left[ 0.0017 \frac{A-2Z}{A} \right]
\\
&Q_{m_\text{e}} = F_\text{A} \left[ 5.5 \cdot 10^{-4} \frac{Z}{A} \right]
\\
&Q_\text{e} = F_\text{A} \left[-1.4 + 8.2 \frac{Z}{A} + 7.7 \frac{Z(Z-1)}{A^{4/3}} \right] \cdot 10^{-4}
\end{align*}
with $Z$ and $N$ the number of protons (atomic number) and neutrons, $A = N+Z$ the nucleon number\footnote{\label{foot:1}Don't confuse the nucleon number $A$ with the index A for a test mass.} and $F_\text{A} := \frac{A \cdot 931 \text{MeV}}{m_\text{A}}$. This last expression can be well approximated by $F_\text{A} \approx 1$, because the mass of the test body $m_\text{A}$ is approximately $A \cdot 931 \text{MeV}$ since the mass of a nucleus minus the average binding energy per nucleus is $931\, \text{MeV}$.
Finally, the dilaton coupling to matter species A is given by
\begin{align*}
\alpha_\text{A}  &= d_\text{g} + (d_{m} - d_\text{g}) Q_{\hat{m}}^\text{A} + (d_{\delta m} - d_\text{g}) Q_{\delta m}^\text{A} + (d_{m_\text{e}} - d_\text{g}) Q_{m_\text{e}}^\text{A} + d_\text{e} Q_\text{e}^\text{A}
\\
&=: d_\text{g} + \bar{\alpha}_\text{A}.
\end{align*}
Notice that while the first term $d_\text{g}$ is matter independent and therefore does not produce any violations of the universality of free fall all other terms depend on the composition of A. Thus they can lead to a measurable E\"{o}tv\"{o}s parameter, providing that the couplings to the individual components of the effective Lagrangian are sufficiently large. Therefore we defined by $\bar{\alpha}_\text{A}$ the part of $\alpha_\text{A}$ which is matter dependent. 
Of course, the dilaton coupling parameters get constrained by null-experiments of the weak equivalence principle.

The potential corresponding to the interaction between two test masses A and B is \cite{Damour.Donoghue}
\begin{align}
\label{equ DIL dilaton interaction potential}
V = -\frac{G m_\text{A} m_\text{B}}{r_\text{AB}}(1 + \alpha_\text{A} \alpha_\text{B} e^{-m_\phi r_\text{AB}}),
\end{align}
where $r_\text{AB}$ is the separation between the masses and we included the possibility that the field has a non-negligible mass $m_\phi$.
In the atmosphere where a test mass A is expected to have a thin shell we therefore have $\alpha_\text{A} \simeq k_{\text{effA}} = 3 \frac{\Delta R_\oplus}{R_\oplus} k$. Remember that the dilaton is assumed to couple with sub-gravitational strength. This matches perfectly with the expectation for a chameleon since there, even if the field couples to matter with about gravitational strength, the effective coupling is much weaker due to the appearance of a bare potential in the Lagrangian.

If we compare the relative acceleration of two test masses A and B in the field of the Earth the E\"otv\"os parameter is
\begin{align*}
\eta_\text{AB} \equiv \left( \frac{\Delta a}{a} \right)_\text{AB} := 2 \frac{a_\text{A} - a_\text{B}}{a_\text{A} + a_\text{B}}
\simeq \alpha_\oplus \left( \alpha_\text{A} - \alpha_\text{B} \right)
= \alpha_\oplus \left( \bar{\alpha}_\text{A} - \bar{\alpha}w_\text{B} \right).
\end{align*}
Assuming that the test masses have a thin shell while the Earth has none, we get $\alpha_\oplus \simeq k_{\text{eff}\oplus} = 3\frac{\Delta R_\oplus}{R_\oplus} k_\oplus$ and $\alpha_\text{A,B} \simeq k_{\text{effA,B}} = k_\text{A,B}$ and therefore
\begin{align*}
\eta_\text{AB} \simeq 3\frac{\Delta R_\oplus}{R_\oplus} k_\oplus \left( k_\text{A} - k_\text{B} \right).
\end{align*}

We conclude that by introducing an appropriate bare potential to the Lagrangian of the dilaton, it might be possible to allow the field to couple to matter with gravitational strength. The presence of the potential would induce a chameleon mechanism, giving an effective coupling $\alpha_\text{A}$ that is much weaker than gravity within the atmosphere. Nevertheless, a proper analysis of the chameleon field which couples to matter as suggested in \cite{Damour.Donoghue} has to be done.

For two given masses A and B the factor $\alpha_\text{A} \alpha_\text{B}$ appearing in the interaction potential \eqref{equ DIL dilaton interaction potential} consists of a linear combination of all possible combinations of the factors $A^{-1/3}$, $\frac{Z}{A}$, $\frac{A-2Z}{A}$,$\frac{(A-2Z)^2}{A^2}$ and $\frac{Z(Z-1)}{A^{4/3}}$ for both test masses A and B. Using $Z(Z-1) \approx Z^2$ and that $A \approx 2Z$ for stable isotopes it is argued that the coefficients appearing in front of the factors $A^{-1/3}$ and $\frac{Z^2}{A^{-4/3}}$ are the dominant contributions to violations of the equivalence principle and that therefore the other terms can be neglected for a good approximation. For this approximation the two dilaton charges are
\begin{align*}
&Q^\prime_{\hat{m}} = - \frac{0.036}{A^{1/3}} - 1.4 \cdot 10^{-4} \frac{Z(Z-1)}{A^{4/3}}
\\
&Q^\prime_\text{e} = + 7.7 \cdot 10^{-4} 	 \frac{Z(Z-1)}{A^{4/3}}
\end{align*}
and therefore the coupling becomes
\begin{align*}
\alpha_\text{A}  &= d_\text{g} + (d_{m} - d_\text{g}) Q_{\hat{m}}^{\prime\text{A}} + d_\text{e} Q_\text{e}^{\prime\text{A}}.
\end{align*}
So what about the coupling parameters $d_\text{a}$? These are the parameters of the theory which can be determined or constrained by testing the universality of free fall. To fully determine those five parameters one would need five independent and different experiments, each detecting a violation of the weak equivalence principle.
In principle they can take any value but it is argued in \cite{Damour.Donoghue} that it is likely to have either $d_\text{e} \sim d_\text{g} - d_{\hat{m}}$ or $d_\text{e} \sim (d_\text{g} - d_{\hat{m}}) / 40$.

Further it is reasonable to assume that $d_\text{g}$ dominates since this is the term responsible for the non-universal part of the coupling which does not lead to violations of the weak equivalence principle. Using this we can calculate the E\"{o}tv\"{o}s parameter for the approximate coupling
\begin{align*}
\eta_\text{AB} &\simeq \alpha_\oplus \left( \bar{\alpha}_\text{A} - \bar{\alpha}w_\text{B} \right)
\\
&= \left( d_\text{g} + (d_{m} - d_\text{g}) Q_{\hat{m}}^{\prime\oplus} + d_\text{e} Q_\text{e}^{\prime\oplus} \right)
\left( (d_{m} - d_\text{g}) (Q_{\hat{m}}^{\prime\text{A}}-Q_{\hat{m}}^{\prime\text{B}}) + d_\text{e} (Q_\text{e}^{\prime\text{A}}-Q_\text{e}^{\prime\text{B}}) \right)
\\
&\simeq d_\text{g} (d_{m} - d_\text{g}) (Q_{\hat{m}}^{\prime\text{A}}-Q_{\hat{m}}^{\prime\text{B}})
+ d_\text{g} d_\text{e} (Q_\text{e}^{\prime\text{A}}-Q_\text{e}^{\prime\text{B}})
\\
&= D_{\hat{m}} (Q_{\hat{m}}^{\prime\text{A}}-Q_{\hat{m}}^{\prime\text{B}})
+ D_\text{e} (Q_\text{e}^{\prime\text{A}}-Q_\text{e}^{\prime\text{B}})
\end{align*}
where we defined
\begin{align*}
&D_{\hat{m}} := d_\text{g} (d_{m} - d_\text{g})
\\
&D_\text{e} :=  d_\text{g} d_\text{e}.
\end{align*}
The constraints on these parameters, given by the E\"{o}t-Wash experiment and lunar laser ranging, respectively, are \cite{Damour.Donoghue.short}
\begin{align}
\begin{split}
&\left| D_{\hat{m}} + 0.22 D_\text{e} \right| \le 5.1\cdot 10^{-11}
\\
&\left| D_{\hat{m}} + 0.28 D_\text{e} \right| \le 24.6 \cdot 10^{-11}.
\end{split}
\label{equ DIL current constraints of approx model}
\end{align}
Notice that $d_\text{g}$ can be constrained by limits on the PPN parameter $\gamma$ since \cite{Damour.Donoghue}
\begin{align*}
d_\text{g}^2 \simeq \frac{1-\gamma}{2}.
\end{align*}

\subsection{Potential Dilaton Constraints for STE-QUEST and MICROSCOPE}
In this section we will discuss what the two satellite missions STE-QUEST and MICROSCOPE can find out about dilaton scalar fields, as was already done partially in \cite{Damour.Donoghue}.

According to \cite{Damour.Vokrouhlicky}, the Earth consists of a core made of iron (Fe), contributing $32\%$ to the total Earth mass, and a mantle made of silicon dioxide ($\text{SiO}_2$), contributing the remaining $78\%$ to the mass of the Earth. 
For the most abundant isotope of iron, $Z = 26$, $N = 30$, $A = 56$ and $m_\text{Fe} = 52'103.06\,\text{MeV}$. A silicon dioxide molecule has an atomic number of $Z = 14 + 2\cdot 8 = 30$ and therefore these two species are not too distinct in terms of their dilaton charges.
%For $\text{SiO}_2$ we have $Z = 14 + 2 \cdot 8 = 30$, $N = 30$ and therefore $A = N + Z = 60$ since both silicon Si and oxygen O have their main abundance with $Z = N$ and further we have $m_{\text{SiO}_2} = 55'968.2\,\text{MeV}$. 
Because of that we consider the Earth to consist of iron only for simplicity, which has the dilaton charges given in table \ref{tab DIL dilaton charges Fe}.
\begin{table}[H]
\centering
\begin{tabular}[c]{l|ll|llll}
 & $Z$ & $A$ & $Q_{\hat{m}}$ & $Q_{\delta m}$ & $Q_{m_\text{e}}$ & $Q_e$\\ \hline
Iron $^{56}\text{Fe}$ & $26$ & $56$ & $8.312 \cdot 10^{-2}$ & $0.012 \cdot 10^{-2}$ & $0.026 \cdot 10^{-2}$ & $0.258 \cdot 10^{-2}$ \\ \hline
 \end{tabular}
\caption{Dilaton charges for iron Fe.}
\label{tab DIL dilaton charges Fe}
\end{table}
%With this we can calculate the E\"{o}tv\"{o}s parameter
%\begin{align*}
%\eta_{^{87}\text{Rb}^{85}\text{Rb}} &\simeq \alpha_\oplus(\alpha_{^{87}\text{Rb}} - \alpha_{^{85}\text{Rb}})
%\\
%&\simeq d_\text{g} (d_{m} - d_\text{g}) \left( Q_{\hat{m}}^{^{87}\text{Rb}} - Q_{\hat{m}}^{^{85}\text{Rb}} \right)
%+ d_\text{g} (d_{\delta m} - d_\text{g}) \left( Q_{\delta m}^{^{87}\text{Rb}} - Q_{\delta m}^{^{85}\text{Rb}} \right)
%\\
%&\quad+ d_\text{g} (d_{m_\text{e}} - d_\text{g}) \left( Q_{m_\text{e}}^{^{87}\text{Rb}} - Q_{m_\text{e}}^{^{85}\text{Rb}} \right)
%+ d_\text{g} d_\text{e} \left( Q_{m_\text{e}}^{^{87}\text{Rb}} - Q_{\text{e}}^{^{85}\text{Rb}} \right)
%\end{align*}
%which contains all the five parameters $d_\text{a}$. Notice that in the last step we assumed that the parameter $d_\text{g}$ dominates over all others, but in doing so we lost the information about the composition of the source of the external field.
\\
\\
In the differential accelerometer used in MICROSCOPE the test masses are made of titanium and platinum. Since these two species are located at very different positions in the period table it is sufficient to apply the approximation discussed above. There, only two dilaton charges, which are very likely to dominate the weak equivalence principle violation signals, have to be considered (see Table \ref{tab DIL dilaton charges approx}).
\begin{table}[H]
\centering
\begin{tabular}[c]{l|ll|ll}
 & $Z$ & $A$ & $Q^\prime_{\hat{m}}$ & $Q^\prime_{\text{e}}$ \\ \hline
Titanium $^{48}\text{Ti}$& $22$ & $48$ & $-10.28 \cdot 10^{-3}$ & $2.04 \cdot 10^{-3}$ \\
Platinum $^{195}\text{Pt}$ & $78$ & $195$ & $-6.95 \cdot 10^{-3}$ & $4.09 \cdot 10^{-3}$ \\
Rubidium $^{87}\text{Rb}$ & $37$ & $87$ & $-8.61 \cdot 10^{-3}$ & $2.66 \cdot 10^{-3}$ \\
Potassium $^{41}\text{K}$ & $19$ & $41$ & $-10.78 \cdot 10^{-3}$ & $1.86 \cdot 10^{-3}$ \\
Iron $^{56}\text{Fe}$ & $26$ & $56$ & $-9.83 \cdot 10^{-3}$ & $2.34 \cdot 10^{-3}$ \\ \hline
\end{tabular}
\caption{Approximate dilaton charges for titanium Ti, platinum Pt, rubidium Rb, potassium K and iron Fe.}
\label{tab DIL dilaton charges approx}
\end{table}
So we can calculate
\begin{align*}
\vec{Q}_{\text{Ti}\text{Pt}} &:= (Q_{\hat{m}}^{\prime\text{Ti}} - Q_{\hat{m}}^{\prime\text{Pt}}, Q_{e}^{\prime\text{Ti}} - Q_{e}^{\prime\text{Pt}})
\\
&= (-3.32, -2.05) \cdot 10^{-3}
\end{align*}
For the experiment performed on MICROSCOPE we obtain the E\"otv\"os parameter
\begin{align*}
&\eta_\text{TiPt} = \alpha_\oplus(\alpha_{\text{Ti}} - \alpha_{\text{Pt}})
\\
&\simeq D_{\hat{m}} (Q_{\hat{m}}^{\prime\text{Ti}} - Q_{\hat{m}}^{\prime\text{Pt}})
+ D_\text{e} (Q_\text{e}^{\prime\text{Ti}} - Q_\text{e}^{\prime\text{Pt}}).
\\
&= -\left( 3.32 \cdot D_{\hat{m}} + 2.05 \cdot D_\text{e} \right) \cdot 10^{-3}
\end{align*}
and since MICROSCOPE aims to test the E\"{o}tv\"{o}s parameter down to $\eta < 10^{-15}$ it can give the constraint
\begin{align*}
\left| D_{\hat{m}} + 0.62 D_\text{e} \right| \le 3.01 \cdot 10^{-13}.
\end{align*}
MICROSCOPE will be able to improve the constraints on the two parameters $D_{\hat{m}}$ and $D_\text{e}$ of the approximate model about two orders of magnitude compared to the current constraints \eqref{equ DIL current constraints of approx model}.

The satellite mission STE-QUEST aims to perform a quantum test of the weak equivalence principle by comparing the propagation of matter wave of two different species: rubidium $^{87}$Rb and potassium $^{41}$K (see Table \ref{tab DIL dilaton charges approx}). Taking into accound the goal of $\eta < 2 \times 10^{-15}$, we obtain the constraint
\begin{align*}
\left| D_{\hat{m}} + 0.37 D_\text{e} \right| \le 9.21 \cdot 10^{-13}.
\end{align*}

\section{Conclusions}
There has been dramatic improvement in frequency standards over the past decade. More and more precise atomic clocks are being placed in space and global 
atomic clock networks are being planned on Earth, which communicate with the clocks in space. The fiber links used to connect the clocks on Earth are often also
 used for internet, and provide a communication speed that is no longer limited by the network, but instead by our computers, our storage devices,
 and our ability to interact with this information. It is important to consider potential applications arising from this amazing technology both in space and on Earth.

Already at the available stability of $\Delta f/f \sim 10^{-16}$, a space mission like the Redshift explorer could reconstruct the space-craft's 4D trajectory from
the observed tick rate of the clock. The limitations currently come from the link between the space clock and the best available clocks on the ground. Such an
experiment could test a multitude of higher order relativistic effects in the same experimental setting. The details were presented in Fig. \ref{fig-2}. Furthermore,
weak field general relativity may potentially be constrained around the Earth via PPN parameters to the $|\gamma- 1| \sim |\beta -1| \sim 10^{-6}$ level (Fig.\ \ref{fig-4}).

 We also show that MICROSCOPE and STE-QUEST will be able to improve current constraints on the dilaton, given by E\"{o}t-Wash experiments and 
 lunar laser ranging, by about two orders of magnitude. The experiments are different in that the MICROSCOPE experiment tests the universality of free
 fall for macroscopic test masses, while the atom interferometer on board STE-QUEST performs a similar experiment at the quantum level.

\end{document}